\documentclass[ApJL,twocolumn]{aastex62}
\usepackage{amssymb, amsmath, graphicx, dcolumn, color, units, xspace, mathtools, physics, tensor, bm, lipsum, revsymb}
\usepackage{lineno}
\usepackage{subfigure}
\usepackage[normalem]{ulem}
\usepackage{import} 
\usepackage{multirow}
\usepackage{xspace}
\usepackage{acronym}
\usepackage{xcolor}
\colorlet{RED}{red}

\definecolor{ForestGreen}{RGB}{34,139,34}
\definecolor{Awesome}{rgb}{1.0, 0.13, 0.32}

\newcommand{\SPA}{School of Physics and Astronomy, Monash University, Clayton VIC 3800, Australia}
\newcommand{\OzGravMonash}{OzGrav: The ARC Centre of Excellence for Gravitational Wave Discovery, Clayton VIC 3800, Australia}
\newcommand{\OzGravSwin}{OzGrav: The ARC Centre of Excellence for Gravitational Wave Discovery, Hawthorn VIC 3122, Australia}
\newcommand{\Swin}{Centre for Astrophysics and Supercomputing, Swinburne University of Technology, Hawthorn VIC 3122, Australia}
\newcommand\CSIRO{CSIRO, Space and Astronomy, PO Box 76, Epping, NSW 1710, Australia}

\begin{document}
\shorttitle{Toward robust detections of nanohertz gravitational waves}
\shortauthors{Di Marco \textit{et al}.}
\title{Toward robust detections of nanohertz gravitational waves}

\author[0000-0003-3432-0494]{Valentina Di Marco}
\affiliation{\SPA}
\affiliation{\OzGravMonash}
\affiliation{\CSIRO}
\author[0000-0002-9583-2947]{Andrew Zic}
\affiliation{\CSIRO}

\author[0000-0002-5455-3474]{Matthew T. Miles}
\affiliation{\Swin}
\affiliation{\OzGravSwin}
\author[0000-0002-2035-4688]{Daniel J. Reardon}
\affiliation{\Swin}
\affiliation{\OzGravSwin}
\author[0000-0002-4418-3895]{Eric Thrane}
\affiliation{\SPA}
\affiliation{\OzGravMonash}
\author[0000-0002-7285-6348]{Ryan M. Shannon}
\affiliation{\Swin}
\affiliation{\OzGravSwin}

\correspondingauthor{Valentina Di Marco}
\email{valentina.dimarco@monash.edu}

\date{\today}


\begin{abstract}
The recent observation of a common red-noise process in pulsar timing arrays (PTAs) suggests that the detection of nanohertz gravitational waves might be around the corner.
However, in order to confidently attribute this red process to gravitational waves, one must observe the Hellings-Downs curve---the telltale angular correlation function associated with a gravitational-wave background.
This effort is complicated by the complex modelling of pulsar noise.
Without proper care, misspecified noise models can lead to false-positive detections. 
Background estimation using ``quasi-resampling'' methods such as sky scrambles and phase shifts, which use the data to characterize the noise, are therefore important tools for assessing significance.
We investigate the ability of current PTA experiments to estimate their background with ``quasi-independent'' scrambles---characterized by a statistical ``match'' below the fiducial value: $|M|<0.1$.
We show that sky scrambling is affected by ``saturation'' after ${\cal O}(10)$ quasi-independent realizations; subsequent scrambles are no longer quasi-independent.
We show phase scrambling saturates after ${\cal O}(100)$ quasi-independent realizations.
With so few independent scrambles, it is difficult to make reliable statements about the  $\gtrsim 5 \sigma$ tail of the null distribution of the detection statistic.
We sketch out various methods by which one may increase the number of independent scrambles.
We also consider an alternative approach wherein one re-frames the background estimation problem so that the significance is calculated using statistically \textit{dependent} scrambles.
The resulting $p$-value is in principle well-defined but may be susceptible to failure if assumptions about the data are incorrect.
\end{abstract}

\section{Introduction}\label{introduction}
Pulsar timing arrays (PTAs) use the stable pulse arrival-time measurements from millisecond-period pulsars to search for gravitational waves \citep{Foster_1990}.

Passing gravitational waves perturb the space-time metric along the line of sight to each pulsar, changing the proper separation between the pulsar and Earth, which induces deviations in the pulsar times of arrival. 
A key science goal of PTAs is to detect a stochastic gravitational-wave background generated by the incoherent superposition of unresolved signals \citep[e.g.,][]{Jenet_2005, Manchester_2013}.
Pulsar timing arrays are sensitive to nanohertz-frequency gravitational waves that can originate from inspiralling supermassive black holes \citep{Rajagopal_1995, Phinney_2001, Wyithe_2003} as well as more exotic sources like cosmic strings \citep{Damour_2000, Damour_2001, Sanidas_2013}, and phase transitions in the early Universe \citep{Maggiore_2000, Maggiore_2001, Caprini_2010}. 

The stochastic background is likely to be approximately isotropic \citep{Rosado_2012}; however significant effort has been made to develop a methodology for detecting anisotropic backgrounds as well \citep{Mingarelli_2013}.
An isotropic background induces a correlation between pulsar pairs, which depends only on the angle of separation $\theta$ between each pulsar.
The predicted angular correlation function for an isotropic background is known as the Hellings \& Downs curve \citep{Hellings_1983} and has the form:
\begin{equation}
    \Gamma_{ab} (\theta) = \frac{1}{2} - \frac{1}{4}x + \frac{3}{2} x \ln(x),
\end{equation}
where $x = (1 - \cos \theta)/2$.
It is sometimes referred to as an ``overlap reduction function''; see also \cite{Christensen_1992, Finn_2008, Allen_1997}.
Detection of the Hellings \& Downs correlation is considered the definitive signature of a nanohertz gravitational wave background.
To make a detection, therefore, it is necessary to observe and time a large set of pulsars spread across many angles and for many years; see, e.g., \cite{Taylor_2016}.

A number of timing-array experiments globally embraced this challenge, including the European PTA \citep[EPTA; ][]{Kramer_2013}, the Indian PTA \citep[InPTA;][]{Joshi_2018}, the North American Nanohertz observatory for gravitational waves \citep[NANOGrav;][]{McLaughlin_2013}, and the Parkes PTA \citep[PPTA;][]{Manchester_2013}.
Together, these form the International PTA \citep[IPTA;][]{Hobbs_2010}.
Other recent PTAs include the Chinese PTA \citep{LeiQian_2016, Lee_2016} and the MeerKAT PTA \citep{bailes_2020,miles_2023}. 
The sensitivity of a PTA to gravitational waves improves with the addition of new pulsars into the arrays, as well as with enhancements in timing precision and improvements in data-analysis techniques \citep{Siemens_2013}. 
With these improvements, \cite{Taylor_2016} have suggested that the stochastic gravitational wave background is expected to be detected this decade.

Indeed, there is mounting support for the existence of a gravitational-wave background. 
The first sign of a possible signal came in 2020 with the observation of a common-spectrum red noise process consistent with a stochastic gravitational-wave background, first by NANOGrav \citep{Arzoumanian_2020}, and subsequently by PPTA \citep{Goncharov_2021A}, EPTA \citep{Chen_2021}, and the IPTA \citep{Antoniadis_2022}. 
More recently, evidence of a Hellings \& Downs correlation has been reported by several PTAs.
NANOGrav observes a signal consistent with a gravitational-wave amplitude $A_\mathrm{GWB} = 6.3^{+4.1}_{-2.5} \times 10^{-15}$ and spectral index $\gamma = 3.2^{+0.6}_{-0.5}$ (90\% credible intervals; $p$-value = $\lesssim 10^{-4}$ \cite{Agazie_2023}).
The CPTA observes a signal with $\log_{10} A_\mathrm{GWB} = -14.4^{+1.0}_{-2.8}$ and $\gamma \in (-1.8, 1.5)$ (95\% credibility, $p$-value = $4 \times 10^{-6}$; \cite{Xu_2023}).
Using a subset of their data, the EPTA observes a signal with $\log_{10} A_\mathrm{GWB} = -13.94^{+0.23}_{-0.48}$ and $\gamma = 2.71^{+1.18}_{-0.73}$ (90\% credibility; $p$-value $\lesssim 0.001$; \cite{Antoniadis_2023}).
The PPTA dataset, meanwhile, yields only a marginal detection ($p$-value $<0.02$), with $\log_{10} A_\mathrm{GWB} = -14.51_{-0.2}^{+0.18}$ and $\gamma = -3.87 \pm 0.47$ (95\% credibility).
We return to these latest results, which were made public after the submission of this manuscript, in Section~\ref{epilogue}.

Confident detection of the Hellings \& Downs curve is not straightforward.
Since it is difficult to quantify the systematic errors in pulsar noise models, ``quasi-resampling'' methods have been developed to assess the significance of the Hellings \& Downs correlation \citep{Cornish_2016, Taylor_2017}.\footnote{We define quasi-resampling as the practice of using resampled, modified data to empirically estimate the significance of a detection statistic. It is similar in spirit to various resampling procedures such as bootstrapping and permutation~\citep{Efron_1993}, except part of the data is overwritten with new (often simulated) values in order to remove some feature from the data (usually the signal) while preserving the salient features of the noise.}
The $p$-value is the probability of observing some detection statistic in excess of the observed value given the null hypothesis.  For our purpose, the null hypothesis is that there is no gravitational-wave signal present in the data.
There are two well-established forms of quasi-resampling in PTA experiments.
\textit{Sky scrambling} creates noise realisations by randomly re-assigning the sky locations of each pulsar, thereby removing any Hellings \& Downs correlations present in the data (but preserves signals whose degree of correlation depends weakly on changes in position, e.g. a monopole).
In contrast, \textit{phase scrambling} (also referred to as ``phase shifting'') preserves the pulsar locations but randomly shifts the residual arrival times when measuring cross-correlations, which can also remove correlations, regardless of their angular dependence. 
Since quasi-resampling noise models are data-driven, they are more robust to the systematic error of pulsar noise models.
However, care must be taken to ensure that quasi-resampling accurately reflects the underlying distribution of the measurement statistic.

In this paper, we examine how pulsar timing arrays estimate the significance of candidate gravitational-wave signals using scrambling methods.
We show how sky scrambling and phase scrambling both suffer from ``saturation,'' which limits the number of independent noise realisations that can be generated.
We show that sky scrambling saturates after ${\cal O}(10)$ quasi-independent noise realisations while phase scrambling saturates after ${\cal O}(100)$ quasi-independent noise realisations.
We explain how this likely limits our ability to understand the null distribution of the detection statistic.
We sketch out different means by which PTAs can increase the number of quasi-independent scrambles.
We also discuss an alternative approach, which employs statistically \textit{dependent} scrambles.
We suggest that the two approaches test two different hypotheses, but they both yield well-defined $p$-values, which can be used to falsify the null hypothesis that the data are well described by some noise model.

The remainder of this paper is outlined as follows.
In Section~\ref{detection_and_mis-specification} we review the basics of gravitational-wave detection. 
In Section~\ref{quasi-independent}, we examine the notion of quasi-independent noise realizations and derive metrics for measuring the dependence of different scrambles.
In Section~\ref{bootstrap}, we show that the two common methods of sky scrambles and phase scrambles saturate for current PTAs after ${\cal O}(10-100)$ quasi-independent noise realisations.\footnote{The assumptions underpinning this result, including our requirement on the ``match'' statistic, are described in detail below.}
We explain how this limits our knowledge of the null distribution for the detection statistic.
In Section~\ref{other} we discuss possible solutions including the use of statistically dependent scrambles.
We also discuss strategies for generating additional independent scrambles.
Finally, in Section~\ref{discussion}, we provide an overview of the key points from this paper and highlight interesting questions for future study.

\section{Detection \& misspecification}\label{detection_and_mis-specification}

\subsection{Detection Basics}
In pulsar timing, an isotropic Gaussian  gravitational-wave background can be characterized by its characteristic strain spectrum, which is frequently assumed to follow a power law \citep{Rajagopal_1995, Phinney_2001, Wyithe_2003, Jaffe_2003}
\begin{align}
    h_c(f) =  A_{\alpha} \left(\frac{f}{f_\text{ref}}\right)^\alpha.
\end{align}
Here, $f_{\text{ref}} = \unit[1]{yr^{-1}}$ is a  reference frequency, $A_\alpha$ is the amplitude of the signal and $\alpha$ is the spectral index.
In the case of binary black holes driven to merge through gravitational-wave emission $\alpha=-2/3$ \cite[][]{Phinney_2001}.
A stochastic background is detected when $A_{\alpha}$ is shown to be inconsistent with zero.

Pulsar-timing searches for the stochastic background rely on the principle of cross-correlation.
While various Bayesian detection statistics have been proposed, see, e.g., \cite{Becsy_2021}, for our purposes here, it is convenient to focus on a frequentist optimal statistic \citep{Allen_1997, Anholm_2009, Chamberlin_2015}, which is the easiest to explain.
The optimal signal-to-noise is given by \citep{Allen_1997}:
\begin{align}\label{eq:rho}
    \rho = \frac{\sum_{i\neq j,\mu} s_i^*(f_\mu) s_j(f_\mu) \, Q_{ij}(f_\mu)}
    {\left(\sum_{i\neq j, \mu} Q_{ij}^2(f_\mu) \, P_i(f_\mu) P_j(f_\mu)\right)^{1/2} } .
\end{align}
Here, $s_i$ and $s_j$ are the measured gravitational-wave strains a Fourier decomposition of the arrival times for pulsars $i$ and $j$, which are a function of frequency $f_\mu$.
From hereon, $\mu$ is the $\mu^{\rm{th}}$ frequency of a frequency series.
The quantities $P_i(f_\mu)$, $P_j(f_\mu)$ are the estimated noise power spectral densities for pulsars $i$ and $j$.
Meanwhile, $Q(f_\mu)$ is the optimal filter, which is given by \citep{Allen_1997}:
\begin{align}
    Q_{ij}(f_\mu) \propto & \frac{\Gamma_{ij} \, \Omega_\text{gw}(f_\mu)}
    {f_\mu^3 P_i(f_\mu) P_j(f_\mu)} \\
    = & \frac{\Gamma_{ij} \, S_h(f_\mu)}
    {P_i(f_\mu) P_j(f_\mu)} .
\end{align}
Here, $\Gamma_{ij}$ is the overlap reduction function (the Hellings-Downs curve) for the pulsar pair $ij$ and $\Omega_\text{gw}(f)$ is the dimensionless energy density spectrum for the modeled source.\footnote{The dimensionless energy density is related to the characteristic strain:
\begin{align}
    \Omega_\text{gw}(f) = \frac{2\pi^2}{3H_0^2} f^2 h_c^2(f) .
\end{align}
See \cite{Thrane_2013} for additional details.}
The variable $S_h(f_\mu)$ is the signal power spectral density.

If the signal and noise model are well-specified---and if no signal is present---then the signal-to-noise ratio defined in Eq.~\ref{eq:rho} is expected to be approximately distributed according to a generalized $\chi^2$ with zero mean and unit variance so that a measurement of $\rho \gtrsim 5$ would constitute a convincing detection \citep{Hazboun_2023}.
However, since the noise present in pulsar timing residuals can comprise of many poorly characterized processes \citep{Lentati_2016, Goncharov_2021B}, the noise model may be misspecified, and so the  distribution of $\rho$ under the null hypothesis (when no signal is present) may have different properties (e.g., a different shape or a different width) than what is predicted if the noise models are adequately specified.

\subsection{Estimating significance with scrambling: current practice}
In the context of pulsar timing, quasi-resampling methods are used to estimate the null distribution of a detection statistic, such as $\rho$, using the data to generate empirical realizations of noise.
Quasi-resampling methods are commonplace in astronomy.
For example, LIGO--Virgo--KAGRA frequently use ``time slides'' to estimate the null distribution of their detection statistics \citep{Was_2009, Usman_2016, Capano_2017}.
It is helpful to consider the case of time slides in some detail.

At least two gravitational-wave observatories are required to perform time slides.
The data from one observatory is shifted in time with respect to the data from the other observatory.
If the time-shift is sufficiently large compared to the coherence time of the matched-filter templates used to detect transient gravitational waves, any true signal will no longer be coherent within the two time series.
However, the shifted data preserve key features of the detector noise, which may not be modelled by the idealized Gaussian likelihood function.
These include transient noise artefacts known as ``glitches'' as low-level non-Gaussianity \citep{Blackburn_2008, Powell_2018, Cabero_2019}.
In this way, the shifted data can be used as a realistic representation of the true detector noise.
If the two data streams are now shifted again by an amount that is long compared to the coherence time of the templates, a second \textit{independent} noise realization can be generated.

Repeating this procedure many times can be used to produce a large suite of noise realizations:
\begin{align}
    N_\text{time slides} = T_\text{obs} / t_\text{coh} ,
\end{align}
given by the observation time $T_\text{obs}$ divided by the (maximum) coherence time of the templates $t_\text{coh}$.
In this way, LIGO--Virgo were able to generate $N_\text{time slides} \sim 10^7$ independent noise realizations to support the detection of GW150914 \citep{Abbott_2016}---the first direct detection of gravitational waves and the first discovery of a binary black hole.
Time slides are not a {\em panacea}---all quasi-resampling methods break down at some point; see e.g., \cite{Was_2009,Ashton_2019}.
However, they are often used as the definitive statistic for background estimation in gravitational-wave astronomy; see, e.g., \cite{Abbott_2016,gwtc-1}.

In pulsar timing, sky scrambles and phase shifts \citep{Cornish_2016,Taylor_2017} are two commonly used quasi-resampling methods used for estimating significance.
For sky scrambling, each pulsar is assigned a random sky location $\hat{n}$.
This changes the value of the overlap reduction function in Eq.~\ref{eq:rho}:
\begin{align}\label{eq:sky_scramble}
    \Gamma_{ij}(\hat{n}_i, \hat{n}_j) \rightarrow
    \Gamma_{ij}(\hat{n}_i', \hat{n}_j') .
\end{align}
This in turn spoils any Hellings-Downs correlation that may be present in the data while preserving various noise properties that may not be captured by the likelihood function.
In phase scrambling, the Fourier series of each pulsar is multiplied by a random, frequency-dependent phase
\begin{align}\label{eq:phase_scramble}
    s_i(f_\mu) \rightarrow e^{i \phi_{i,\mu}} s_i(f_\mu) ,
\end{align}
which also has the effect of spoiling any Hellings \& Downs correlation present in the data.

However, the data are manipulated---either with sky scrambles or phase scrambles---the resulting scrambled data can be treated as a realization of realistic detector noise.
Many such realizations are created, and each one is analyzed in order to determine the detection statistic $\rho$ (Eq.\ref{eq:rho}).
In this way, one is able to construct a histogram of $\rho$, which is an empirical estimate for the distribution of $\rho$ under the null hypothesis that no signal is present: $P(\rho | A_\alpha=0)$.
If the unscrambled data yield an optimal signal-to-noise ratio of $\rho_0$, then the associated $p$-value is\footnote{Formally, the distribution $P(\rho | A_\alpha = 0)$ is conditioned on the measured auto-power (comparing a pulsar to itself), which is the same for every scramble. 
Thus, in this framework, we are implicitly assuming that the scrambling estimate of $P(\rho|A_\alpha=0)$ (conditioned on the observed auto-power) is a conservative estimate for the more general distribution (not conditioned on the auto-power).}
\begin{align}\label{eq:pval}
    p = & \int^{+\infty}_{\rho_0} 
    d\rho \, P(\rho | A_\alpha=0) \\
    \approx & \frac{N_\text{scrambles}(\rho > \rho_0)}{N_\text{scrambles}} .
\end{align}

All quasi-resampling methods at some stage ``saturate,'' meaning that there is a limited number of \textit{statistically independent} noise realizations that one can generate; this is true for time slides, phase shifts, sky scrambles, etc.
Increasing the number of realizations beyond this point does not yield independent realisations.
As a result, there is a minimum $p$-value equal to $1/N_\text{scrambles}$ that can be probed with scrambling methods---at least given the way we have set up the problem to this point.
One can attempt to \textit{extrapolate} to smaller $p$-values, or one can estimate $P(\rho | A_{\alpha}=0)$ using a theoretical noise model, but scrambling methods are not able to determine anything more than the fact that $p \lesssim 1/N_\text{scrambles}$ (again, as the problem is currently posed).

In order to ensure that background estimates obtained with scrambling are not affected by saturation in PTA analyses, a match statistic \citep{Taylor_2017,Cornish_2016} is employed to estimate the degree to which sky scrambles are statistically independent:\footnote{A warning to readers skimming this paper for a handy formula: we argue below that this is \textit{not} actually a suitable definition of match.}
\begin{align}\label{eq:match}
    M = \frac{\sum_{i \neq j} \Gamma_{ij} \Gamma_{ij}'}
    {\sqrt{
    \left(\sum_{i \neq j} \Gamma_{ij}^2 \right) 
    \left(\sum_{i \neq j} \Gamma_{ij}'^2  \right) 
    }} .
\end{align}
Here, $\Gamma_{ij}$ and $\Gamma_{ij}'$ are the Hellings-Downs parameters for pulsars $ij$ in two different skies.
When, $\Gamma_{ij} = \Gamma_{ij}'$, the two skies are identical, and so the match is unity.
However, when the skies are different, the match statistic can yield values close to zero, which is interpreted to mean that these two scrambles are quasi-independent.

The established convention in PTA analyses is the requirement that all pairs of scrambles produce a match
\begin{align}\label{eq:threshold}
    |M| < 0.1 .
\end{align}
to ensure quasi-independence \citep{Arzoumanian_2020}, though there is some variation in this threshold value in the literature \citep[e.g.][]{Taylor_2017}.
For the time being, we adopt a threshold of 0.1 as a fiducial value and revisit this choice below.
The match between each scrambled sky and the unscrambled sky is also required to be below this threshold.
If it is not, the scrambled data may include significant contamination from the signal, which makes it harder than needs be to detect a gravitational-wave background.

To the best of our knowledge, the literature does not contain an expression for the match statistic comparing two phase scrambles.
In the next section, we discuss the theoretical framework that underpins these scrambling procedures.
We propose how commonly used metrics for assessing the independence of different scrambling noise realizations can be improved.

\section{Quasi-independent scrambles}\label{quasi-independent}
In this section, we make three main points:
\begin{enumerate}
    \item The commonly used definition of match (Eq.~\ref{eq:match}) is unsuitable for use with real pulsar timing arrays because it does not take into account the relative quality of different pulsars.
    \item Phase scrambles are not automatically quasi-independent.
    One must employ a match statistic analogous to Eq.~\ref{eq:match} in order to determine the extent to which two phase scrambles are independent.
    There are a finite number of quasi-independent phase scrambles.
    \item Sky scrambles and phase scrambles can be combined to generate what we call ``super scrambles.''
    A dedicated match statistic quantifies the statistical independence of two super scrambles.
    There are more quasi-independent super scrambles than there are sky scrambles or phase scrambles.
\end{enumerate}
Each one of these points is associated with a different subsection below.

\subsection{Sky scrambles}
The match can be derived by calculating the covariance for two different sky scrambles:
\begin{align}\label{eq:match_definition}
    M \equiv \langle \rho \rho' \rangle .
\end{align}
Here, $\rho$ is the signal-to-noise ratio associated with one sky scramble and $\rho'$ is the signal-to-noise ratio associated with a different sky scramble.
The angled brackets denote an ensemble average over noise model realizations.
By requiring that the match is close to zero, the two sky scrambles must be approximately independent (on average).
Substituting our expression for $\rho$ (Eq.~\ref{eq:rho}) into the definition of match in Eq.~\ref{eq:match_definition}, we obtain:
\begin{align}
    M = & \frac{
    \sum_{i\neq j, \mu} \frac{\Gamma_{ij}\Gamma_{ij}' \Omega_\text{gw}^2(f_\mu)}
    {f_\mu^6 P_i(f_\mu) P_j(f_\mu)}
    }{ 
    \sqrt{
    \left(\sum_{i\neq j, \mu} \frac{\Gamma_{ij}^2 \Omega_\text{gw}^2(f_\mu)}
    {f_\mu^6 P_i(f_\mu) P_j(f_\mu)} \right)
    \left(\sum_{k\neq l, \mu}  \frac{\Gamma_{kl}'^2 \, \Omega_\text{gw}^2(f_\mu)}
    {f_\mu^6 P_k(f_\mu) P_l(f_\mu)} \right)
    }} \\
    \label{eq:sky_match}
    = & \frac{
    \sum_{i\neq j} \Gamma_{ij}\Gamma_{ij}' w_{ij}
    }{ 
    \sqrt{
    \left(\sum_{i\neq j} \Gamma_{ij}^2 w_{ij} \right)
    \left(\sum_{k\neq l}  \Gamma_{kl}'^2 w_{ij} \right)
    }}
\end{align}
A full derivation is provided in Appendix~\ref{app:sky_match}.\footnote{While the Eq.~\ref{eq:match} definition of match is commonly used in pulsar timing papers, we note that Eq.~\ref{eq:sky_match} is hinted at by Eq.~17 in \cite{Taylor_2017}. A version of match that takes into account noise weighting appears as Eq.~7 in \cite{Cornish_2016}.}
Comparing Eq.~\ref{eq:sky_match} with the commonly used definition of match (Eq.~\ref{eq:match}), it is apparent that each term in the sums of Eq.~\ref{eq:sky_match} is weighted by a weight factor
\begin{align}
    w_{ij} \equiv \sum_\mu 
    \frac{\Omega_\text{gw}^2(f_\mu)}
    {f_\mu^6 P_i(f_\mu) P_j(f_\mu)} 
\end{align}
while Eq.~\ref{eq:match} implicitly assumes that each pulsar pair enters with equal weight.

The weighting factor takes into account the relative importance of each pulsar pair in the optimal signal-to-noise ratio.
Pulsar pairs with relatively lower power spectral densities are weighted with relatively more importance because they provide more signal-to-noise ratio than comparatively noisy pulsar pairs.
The weight also depends on the signal model $\Omega_\text{gw}(f)$, which is proportional to $f^{-2/3}$ for inspiralling black holes.

The presence of weighting factors in Eq.~\ref{eq:sky_match} is intuitive.
If one  considers a pulsar timing network comprising two precisely timed pulsars and 98 poorly timed pulsars, it is clear that there are only two \textit{meaningful} pulsars.
In such a network, one of the weight factors would be much larger than all the other weights.
The commonly used match statistic in Eq.~\ref{eq:match} is the limiting case of Eq.~\ref{eq:sky_match} corresponding to pulsar timing arrays with pulsars of identical quality (with identical noise properties).
However, in realistic pulsar timing arrays, some pulsars are usually timed with far better precision than others -- a fact that is encoded within the optimal detection statistic. 
We show below that there are fewer quasi-independent noise realizations possible when the pulsar quality is taken into account.

\subsection{Phase scrambles}
Repeating the calculation from the previous subsection, we calculate the covariance between two different phase scrambles $\langle \rho \rho' \rangle$.
Using the definition of phase scrambles given in Eq.~\ref{eq:phase_scramble}, we obtain the following expression for match:
\begin{widetext}
    \begin{align}\label{eq:phase_match}
    M = & \frac{
    \sum_{i\neq j, \mu} \Gamma_{ij}^2 
    {f_\mu^6 P_i(f_\mu) P_j(f_\mu)}
    \cos\big(\phi_{j,\mu}-\phi_{i,\mu} + \phi'_{j,\mu}-\phi'_{i,\mu}\big)
    }{ 
    \sum_{i\neq j, \mu} \Gamma_{ij}^2 \frac{\Omega_\text{gw}^2(f_\mu)}
    {f_\mu^6 P_i(f_\mu) P_j(f_\mu)}
    }
\end{align}
\end{widetext}
A full derivation is provided in Appendix~\ref{app:phase_match}.
Comparing this expression with the sky-scramble version of match in Eq.~\ref{eq:sky_match}, we see that the $\Gamma \Gamma'$ terms in the numerator---describing two different skies---are gone, replaced by a factor of $\Gamma^2$.
(There is only one true sky when the data are phase scrambled.)
This time, the scrambling is carried out by four phases for each frequency bin: $\phi_{j,\mu}, \phi_{i,\mu}, \phi'_{j,\mu}, \phi'_{i,\mu}$.
The primed phases correspond to one scramble while the unprimed phases correspond to a different scramble.

As Eq.~\ref{eq:sky_match} should be used to check for the statistical independence of sky scrambles, Eq.~\ref{eq:phase_match} should be used to check for the statistical independence of phase scrambles.
Since phase scrambling has many parameters---one phase per frequency bin per pulsar---one expects to find more quasi-independent noise realisations than one can obtain using sky scrambling.
However, since gravitational-wave signals appear in pulsar-timing searches as red (low-frequency) processes, the additional parameters do not provide as much help as one might naively expect. 
The signal-to-noise ratio depends mostly on the lowest few frequency bins \citep{Arzoumanian_2020}.

\subsection{Super scrambles}
The previous two subsections beg two questions: can sky scrambling and phase scrambling be combined, and if so, what is the resulting match statistic?\footnote{The concept of super scrambles seems to be at least implied in \cite{Taylor_2017}; see their Eq.~26 and surrounding discussion.}
In this case, the match statistic is
\begin{widetext}
\begin{align}\label{eq:super_match}
    M = & \frac{
    \sum_{i\neq j, \mu} \Gamma_{ij}\Gamma_{ij}' \frac{\Omega_\text{gw}^2(f_\mu) }
    {f_\mu^6 P_i(f_\mu) P_j(f_\mu)}
    \cos\big(\phi_{j,\mu}-\phi_{i,\mu} + \phi'_{j,\mu}-\phi'_{i,\mu}\big)
    }{ 
    \left(\sum_{i\neq j, \mu} \Gamma_{ij}^2 \frac{\Omega_\text{gw}^2(f_\mu)}
    {f_\mu^6 P_i(f_\mu) P_j(f_\mu)} \right)^{1/2}
    \left(\sum_{i\neq j, \mu} \Gamma_{ij}'^2 \sum_\mu \frac{\Omega_\text{gw}^2(f_\mu)}
    {f_\mu^6 P_i(f_\mu) P_j(f_\mu)} \right)^{1/2}
    } .
\end{align}
\end{widetext}
We do not provide a derivation since this expression follows the pattern established in Appendices \ref{app:sky_match}-\ref{app:phase_match}.
The numerator contains the $\Gamma \Gamma'$ term that arises from comparing two different skies, but it also contains the four frequency-dependent phases.
Since this super scrambling includes more parameters than sky scrambling alone or phase scrambling alone, one expects to find more quasi-independent noise realizations with super scrambling than with either of the previous methods.

\section{Scrambling with current pulsar timing arrays}\label{bootstrap}

Here we estimate the number of quasi-independent noise realizations available to two current pulsar timing arrays for which we have estimates of the relevant pulsar noise properties: NANOGrav and PPTA.
We use a modified version of the publicly available code \texttt{makeskyscrambles.py}.\footnote{\url{https://github.com/ipta/false_alarm_analyses/}}
The code proposes random sky scrambles using Eq.~\ref{eq:sky_scramble} and/or random phase scrambles using Eq.~\ref{eq:phase_scramble}.
For sky scrambles, the scrambled pulsar coordinates are drawn from an isotropic distribution.
For phase scrambles, the random phases are drawn from a uniform distribution.
In either case, the proposed scramble is compared to the unscrambled data and all the previously accepted scrambles.
The proposal is accepted only if the match criteria $|M|<0.1$ with all pairs of scrambles.\footnote{While performing this analysis, we discovered a bug in this code which meant that the match criteria was not actually enforced, except between the unscrambled data and the proposed scramble. However, this bug is now fixed.}
The publicly available version of the code employs the commonly used match statistic defined in Eq.~\ref{eq:match}.
However, we performed tests using the various other definitions of match described above; see Eqs.~\ref{eq:sky_match}, \ref{eq:phase_match}, and \ref{eq:super_match}.

We estimate pulsar noise curves using the results from \citet{Goncharov_2021B} and data from the PPTA second data release \citep{Kerr_2020}, and likewise from the NANOGrav 12.5-yr data set \citep{Arzoumanian_2020}. 
If the code spends more than two hours unsuccessfully searching for a new scramble, it terminates, and we record the number of quasi-independent scrambles.
It is likely that we could obtain some additional scrambles by allowing the code to run for longer.
However, we expect this would change our results only marginally while greatly increasing computational cost.

\begin{figure}[htp]
    \centering
    \subfigure[NANOGrav 12.5]{\includegraphics[width=0.410\textwidth]{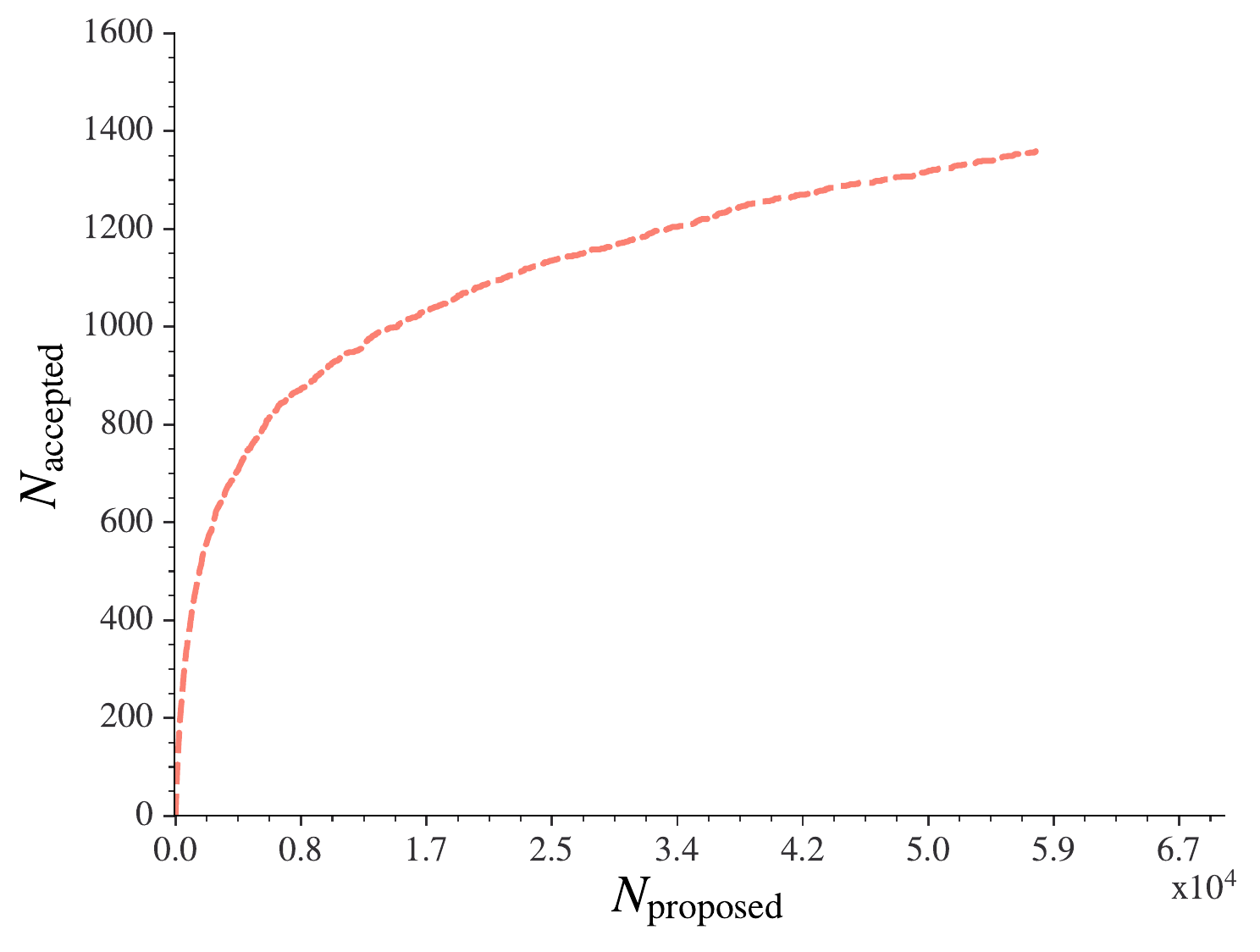}}%
    \qquad
    \subfigure[EPTA]{\includegraphics[width=0.410\textwidth]{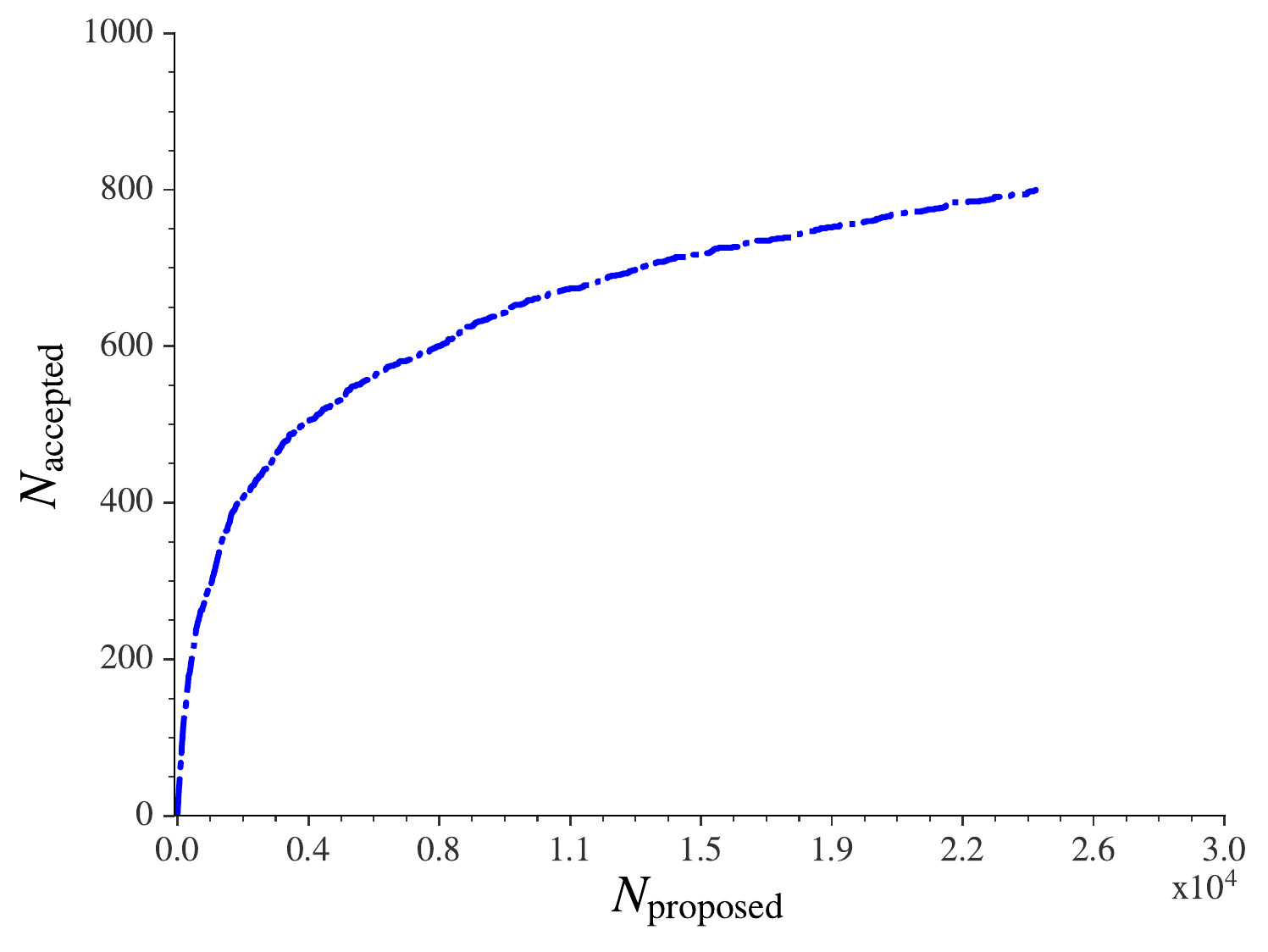}}%
    \qquad
    \subfigure[PPTA]{\includegraphics[width=0.410\textwidth]{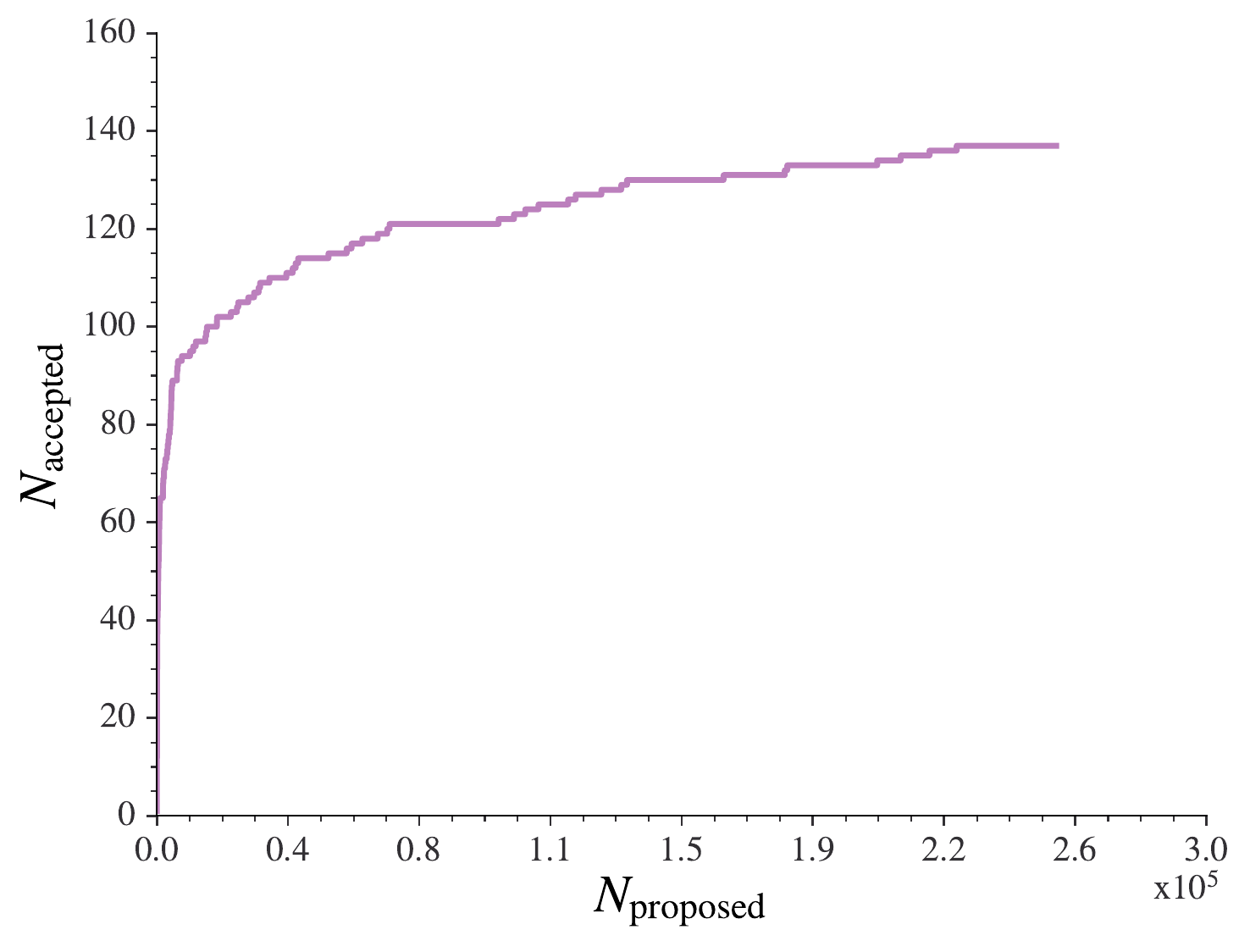}}%
    \caption{
    Saturation plots for sky scrambles assuming equal-quality pulsars (using the match statistic in Eq.~\ref{eq:match}) and requiring $|M|<0.1$.
    In each panel, the vertical axis is the number of accepted scrambles while the horizontal axis is the number of proposed scrambles.
    The top panel (dashed red) shows results for NANOGrav; the middle planel (dot-dash blue) shows results for EPTA; the bottom panel (solid purple) shows results for PPTA.
    We interpret the flattening of each curve as the onset of saturation where it becomes increasingly difficult, and eventually impossible to find new quasi-independent scrambles.
    }
    \label{fig:naive_scramble}
\end{figure}

For our first test, we estimate the number of quasi-independent sky scrambles under the (false) assumption that every pulsar in NANOGrav and PPTA is of equal quality (using the commonly used match statistic defined in Eq.~\ref{eq:match}).
In Fig.~\ref{fig:naive_scramble}, we plot the number of accepted scrambles as a function of the number of proposed scrambles.
For the NANOGrav PTA (top panel, dashed red curve), we find 1,359 quasi-independent sky scrambles.
It is possible one might be able to obtain $\lesssim 1600$ by letting the code run for longer.\footnote{Using a root-finding algorithm, \cite{Bruce} has shown numerically that there are $2N-1$ truly orthogonal scrambles (with $M=0$) in a network of $N$ identical pulsars. We find approximately $2.5$ more than that using a threshold of $|M|<0.1$.}
For the PPTA (bottom panel, solid purple curve), we find that there are 137 quasi-independent scrambles.
In both cases, the code terminates after failing for two hours to find additional scrambles.
Both curves can be seen to asymptote, which we interpret as the beginning of saturation: the point at which it becomes difficult, and eventually impossible to find quasi-independent scrambles.

\begin{figure}[htp]
    \centering
\includegraphics[clip,width=\columnwidth]{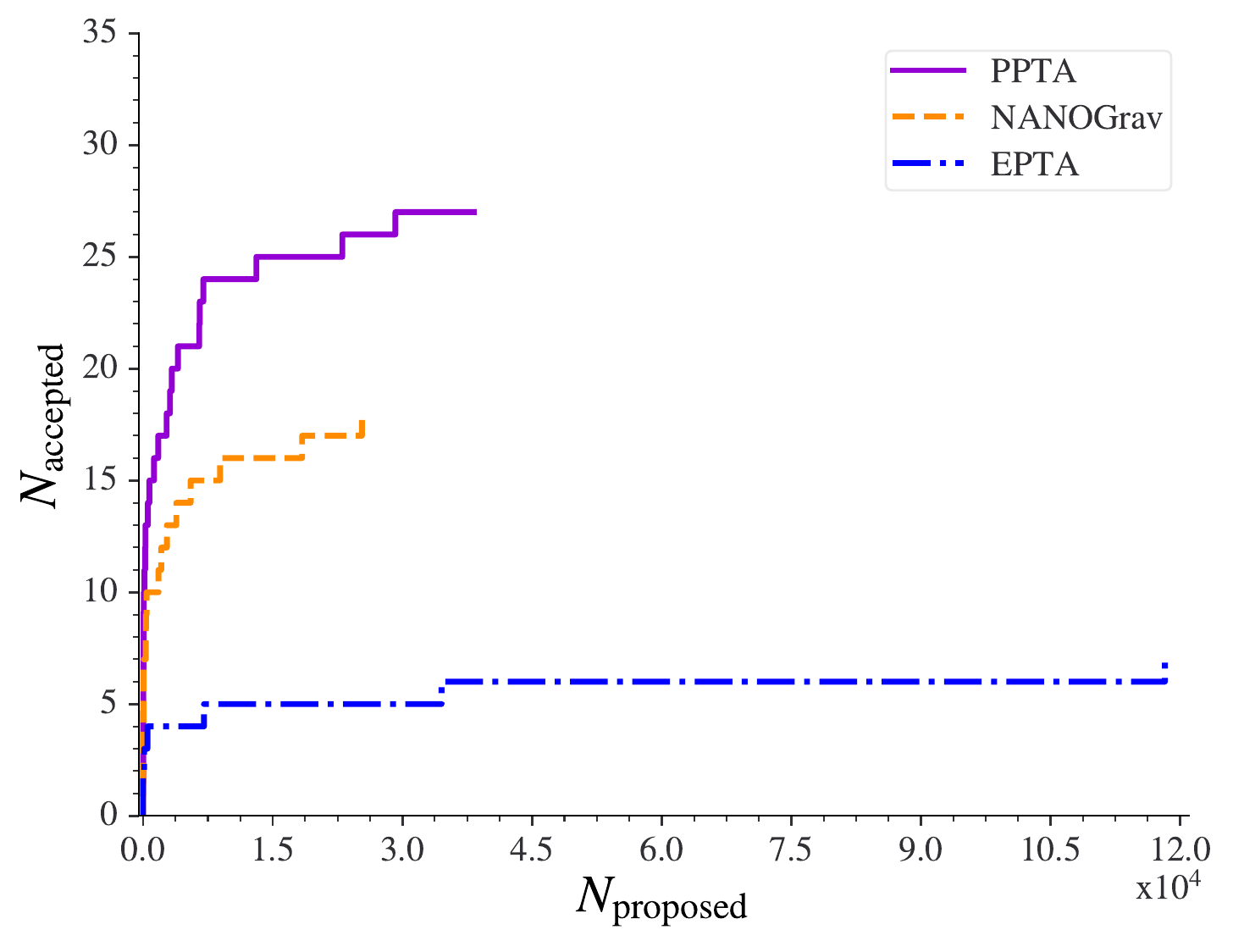}
    \caption{
    Saturation plots for sky scrambles taking into account the relative quality of different pulsars (using the match statistic in Eq.~\ref{eq:sky_match}) and requiring $|M|<0.1$.
    The vertical axis is the number of accepted scrambles while the horizontal axis is the number of proposed scrambles.
    The plot shows results for NANOGrav PPTA and EPTA.
    Comparing with Fig.~\ref{fig:naive_scramble}, we see that saturation occurs much faster when we take into account the quality of different pulsars.
    }
    \label{fig:sky_scramble}
\end{figure}

In Fig.~\ref{fig:sky_scramble}, we include a similar sky scramble ``saturation plot'' except we calculate the match using Eq.~\ref{eq:sky_match} in order to take into account the relative quality of each pulsar.
For the PPTA (solid purple), we find just 27 quasi-independent scrambles while for NANOGrav (dashed orange), we find 18 quasi-independent scrambles.
Comparing Fig.~\ref{fig:sky_scramble} with Fig.~\ref{fig:naive_scramble}, it is evident that the number of quasi-independent noise realisations is dramatically reduced when we take into account the quality of each pulsar.

\begin{figure}[htp]    \includegraphics[clip,width=\columnwidth]{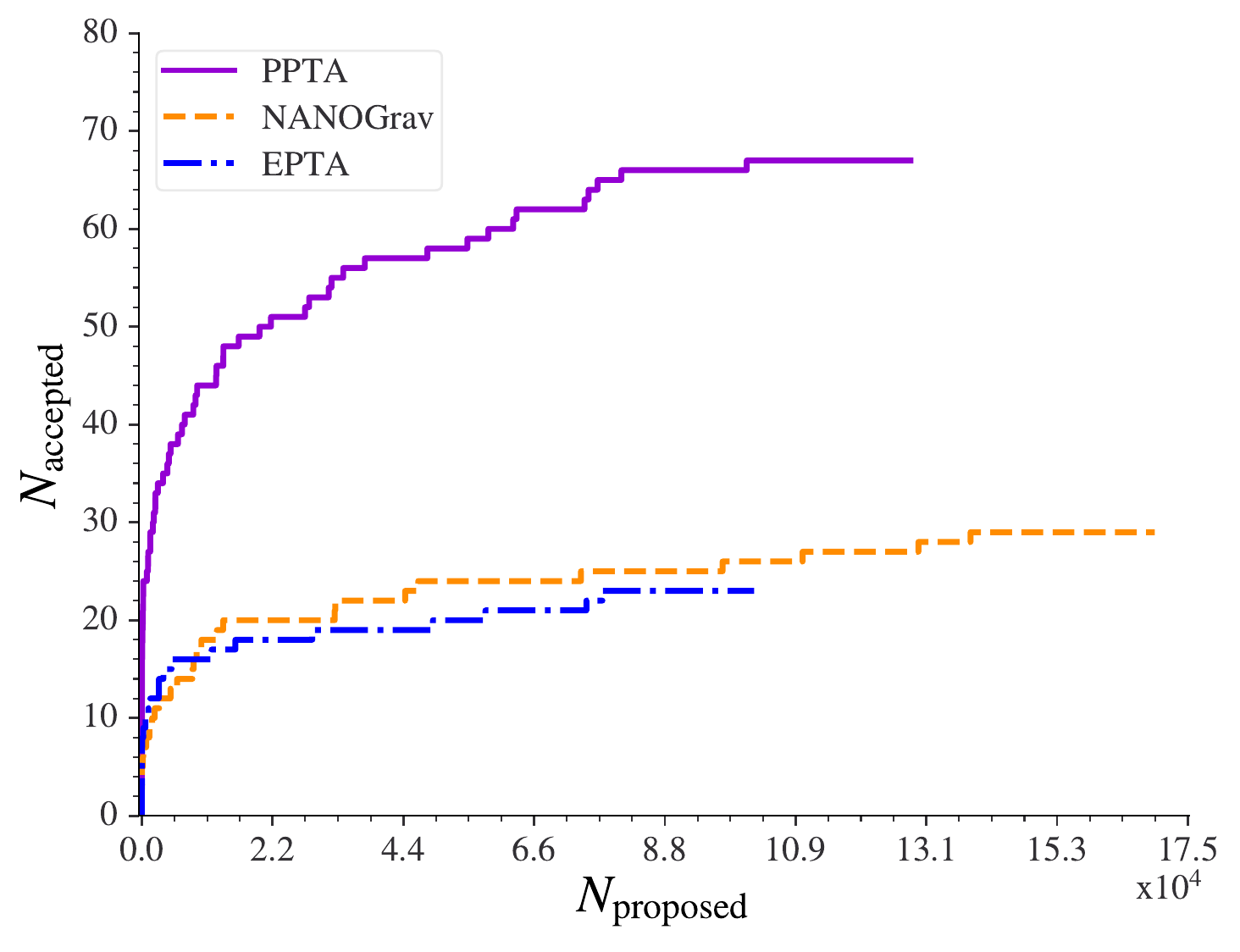}%
    \caption{
    Saturation plots for phase scrambles taking into account the relative quality of different pulsars (using the match statistic in Eq.~\ref{eq:phase_match}) and requiring $|M|<0.1$.
    The vertical axis is the number of accepted scrambles while the horizontal axis is the number of proposed scrambles.
    The plot shows results for NANOGrav PPTA and EPTA.
    Comparing with Fig.~\ref{fig:sky_scramble}, we observe that there are more quasi-independent realisations possible with phase scrambling compared to sky scrambling.
    }
    \label{fig:phase_scramble}
\end{figure}

In Fig.~\ref{fig:phase_scramble}, we present the saturation plot for phase scrambling  using the match defined in Eq.~\ref{eq:phase_match}, which takes into account the relative quality of each pulsar.
For PPTA (solid purple), we find 67 quasi-independent scrambles while for NANOGrav (dashed red), we find 29 quasi-independent scrambles.
Comparing Fig.~\ref{fig:phase_scramble} with Fig.~\ref{fig:sky_scramble}, we observe that phase scrambling provides more quasi-independent scrambles than sky scrambling.

\begin{figure}[htp]
    \centering    \includegraphics[clip,width=\columnwidth]{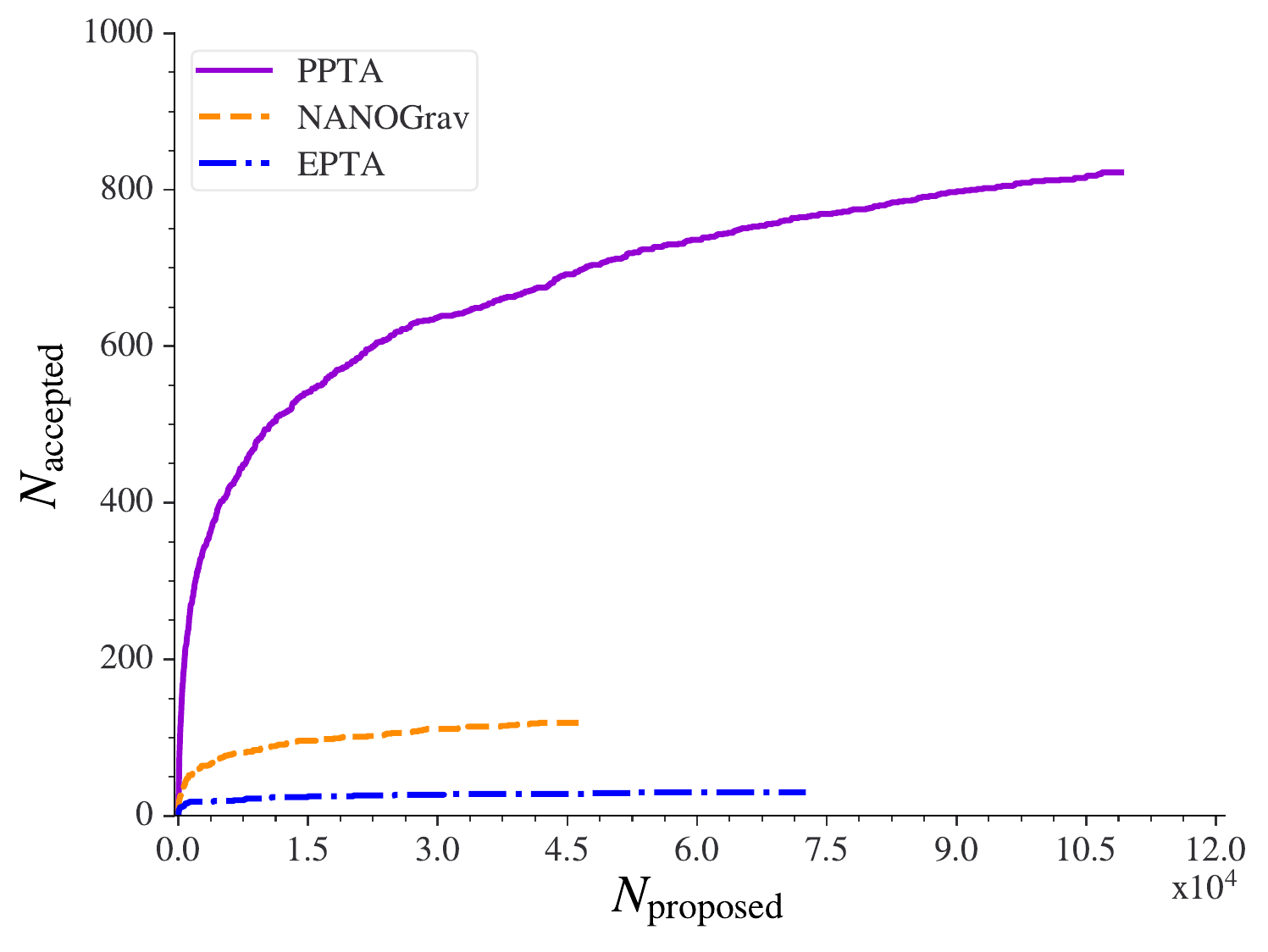}%
    \caption{
    Saturation plot for super scrambles (using the match statistic in Eq.~\ref{eq:super_match}) and requiring $|M|<0.1$.
    The vertical axis is the number of accepted scrambles while the horizontal axis is the number of proposed scrambles.
    The plot shows results for NANOGrav PPTA and EPTA.
    Super scrambling produces more quasi-independent noise realizations than phase scrambling or sky scrambling alone. Note: the NANOGrav curve terminates at a lower $N_\mathrm{proposed}$ due to the two-hour search criterion.}
    \label{fig:super_scramble}
\end{figure}

Finally, in Fig.~\ref{fig:super_scramble}, we present the saturation plot for super scrambling using the match defined in Eq.~\ref{eq:super_match}.
For PPTA (solid purple), we find 822 quasi-independent scrambles while for NANOGrav (dashed red) we find 119 quasi-independent scrambles.
As one would expect, combining phase scrambling with sky scrambling produces more quasi-independent noise realisations than either method by itself.

At this stage, some readers might be puzzled: in Figs.~\ref{fig:sky_scramble}-\ref{fig:super_scramble}, why does NANOGrav, a pulsar timing array known for its large number of pulsars, produce fewer quasi-independent scrambles than the PPTA, which has fewer pulsars?
The answer to this question is that the number of quasi-independent scrambles is not a measure of the overall quality of an array; it is just a measure of the number of influential pulsars that contribute to the signal-to-noise ratio defined in Eq.~\ref{eq:rho}.
Thus, it is possible to add a well-timed pulsar to an existing network and actually \textit{reduce} the number of quasi-independent scrambles.
This does not imply that the network is somehow worse; it just means that the number of influential pulsars has decreased.

Based on these investigations, it seems that the distribution for the detection statistic under the null hypothesis $p(\rho | A=0)$ is not well measured by current pulsar timing arrays.
With only 100-800 quasi-independent noise realizations, it seems impractical to understand the tail of this distribution at the $5\sigma$ level.
In fact, by requiring $|M|<0.1$ we likely overestimate the number of truly independent scrambles (with $M=0$).
As mentioned above, investigations with a root-finding algorithm suggest the number of truly independent scrambles might be lower than our estimates by a factor of $\approx 2.5$ \citep{Bruce}.
In the next section, we discuss strategies to overcome this challenge.

\section{Correlated scrambles and other strategies}\label{other}

\subsection{Detection with correlated scrambles}
Since it seems difficult to estimate the null distribution $p(\rho | A=0)$ with sufficient accuracy to calculate small $p$-values below $1/N_\text{scrambles}$, it appears impractical to falsify with $\gtrsim 5 \sigma$ confidence the original null hypothesis that the unscrambled data are described by $p(\rho | A=0)$.
However, one may choose to re-frame the detection problem so as to avoid this problem.

If one removes entirely the requirement that matches falls below some threshold value as in Eq.~\ref{eq:threshold}, then one can produce an infinite number of statistically \textit{dependent} scrambles with associated $\rho$ values.
Since the associated $\rho$ values are statistically dependent, they are not guaranteed to follow the null distribution $p(\rho | A=0)$; indeed one can concoct examples where these two distributions are different.\footnote{For example, imagine the case where $p(\rho | A=0)$ is a Gaussian distribution, with zero mean and unit variance but the distribution of some number of scrambles all with $M=1$ is a delta function.}
However, they still form \textit{a distribution} that can be used for hypothesis testing so long as there is nothing unique about the unscrambled configuration compared to the scrambled configurations.
A similar argument is put forward by \cite{Usman_2016} in the context of audio-band searches for gravitational waves. The behaviour of LIGO–Virgo $p$-values (calculated with quasi-resampling) was also investigated ahead of the first gravitational-wave detection using a mock data challenge to ensure that, e.g., a $p=0.01$ detection occurs ~1\% of the time \citep{Capano_2017}.

In the next two subsections, we explore this idea in greater detail using simple examples to illustrate the following points:
\begin{enumerate}
    \item In \ref{sound} we show that highly correlated scrambles built on sound assumptions yield well-behaved $p$-values.
    \item In \ref{unsound} we show that highly correlated scrambles built on unsound assumptions can yield false-positive detections.
\end{enumerate}
The remainder of this section sketches out alternative methods for generating additional scrambles should they prove useful.

\subsection{Correlated scrambles with sound assumptions}\label{sound}
This can be illustrated with a pair of examples.
In the first example, we consider a PTA consisting of simply two pulsars that measure the gravitational-wave strain at just one frequency. 
The entire data set consists of just two complex numbers: 
\begin{align}
    h_1 = & A_1 e^{i\phi_1} \\
    h_2 = & A_2 e^{i \phi_2} .
\end{align}
With such a small dataset, we have only a single independent estimate for $\rho$ and so it seems that we know little about the distribution $p(\rho | A=0)$.
Nonetheless, we are free to define the following detection statistic: 
\begin{align}
    X = \cos(\phi_1 - \phi_2) .
\end{align}
If there is a correlated signal present in both pulsars (and the Hellings-Downs curve for this pulsar pair is positive) then $h_1, h_2$ have a tendency to be in phase, which tends to make $X \rightarrow 1$.
If no signal is present in the data, however, we expect $\phi_1, \phi_2$ to be uncorrelated.
Thus, a near-unity value of $X$ can be interpreted as a candidate signal.

One can generate an infinite number of correlated scrambles by drawing random values of $\phi_1, \phi_2$ to build up a distribution for $X$, which can be used to define a $p$-value:
\begin{align}
    p = \frac{N_\text{scrambles}(X>X_0)}
    {N_\text{scrambles}} .
\end{align}
Here, $X_0$ is the detection statistic for the unscrambled data.
It is possible to measure arbitrarily small $p$-values.
These $p$-values are ``correct'' in the sense that a $p=0.001$ observation should occur 0.1\% of the time.
By setting up the problem this way, the hypothesis test is framed directly in terms of our assumptions about the distribution of $\phi_1, \phi_2$.\footnote{There are some curious consequences of this framework.
In particular, consider a network of 26 pulsars like the PPTA DR2 \citep{Kerr_2020}.
To carry out background estimation, we consider a highly contrived method in which we vary the location of just one pulsar in our network.
Moreover, the deviation is restricted to within $1^\circ$ of the pulsar position.
Using this method, it is possible to generate an infinite number of extremely correlated scrambles.
The optimal statistic for each one is very similar in value.
Nonetheless, following the logic of the previous subsections, it should be possible to define a $p$-value.
This is in spite of the fact that we have (intentionally) formulated this method to produce a poor estimate of the null distribution.
One can concoct an analogous thought experiment in which one carries out background estimation for a pair of audio-band gravitational-wave detectors by producing many statistically-dependent time-slides with lags all less than the coherence time of the matched-filter template bank.}

\subsection{Correlated scrambles with unsound assumptions}\label{unsound}
In the previous subsection, we argue that in theory, it is possible to derive reliable $p$-values with only $\approx 1$ independent scramble.
In both cases, this conclusion follows from the assumption that the unscrambled configuration is fungible with the scrambled configurations.
We now ask: can we concoct a scenario where this assumption breaks down?

We consider a network of just two pulsars.
The noise in each pulsar consists of step functions, each with a random rise time and a random sign.
The amplitude is the same for every draw.
We choose this form of noise so that the phases of different frequency bins are correlated, which violates an implicit assumption of the phase scrambling algorithm.
While this noise is not a realistic model for pulsar timing noise, which is probably best described as quasi-Gaussian, there are sometimes step-function-like jumps in pulsar timing residuals from both instrumental artifacts \citep[such as changes to the back end; e.g.][]{Kerr_2020} and astrophysical processes such as pulsar glitches or profile change events \citep[e.g., ][]{Yu_2013,Shannon_2016,Singha_2021,Jennings_2022}.
PTA collaborations endeavour to characterize and remove such jumps, but it is possible that some low-level offsets persist, creating non-stationary artefacts in the noise.

\begin{figure}[htp]    \includegraphics[clip,width=\columnwidth]{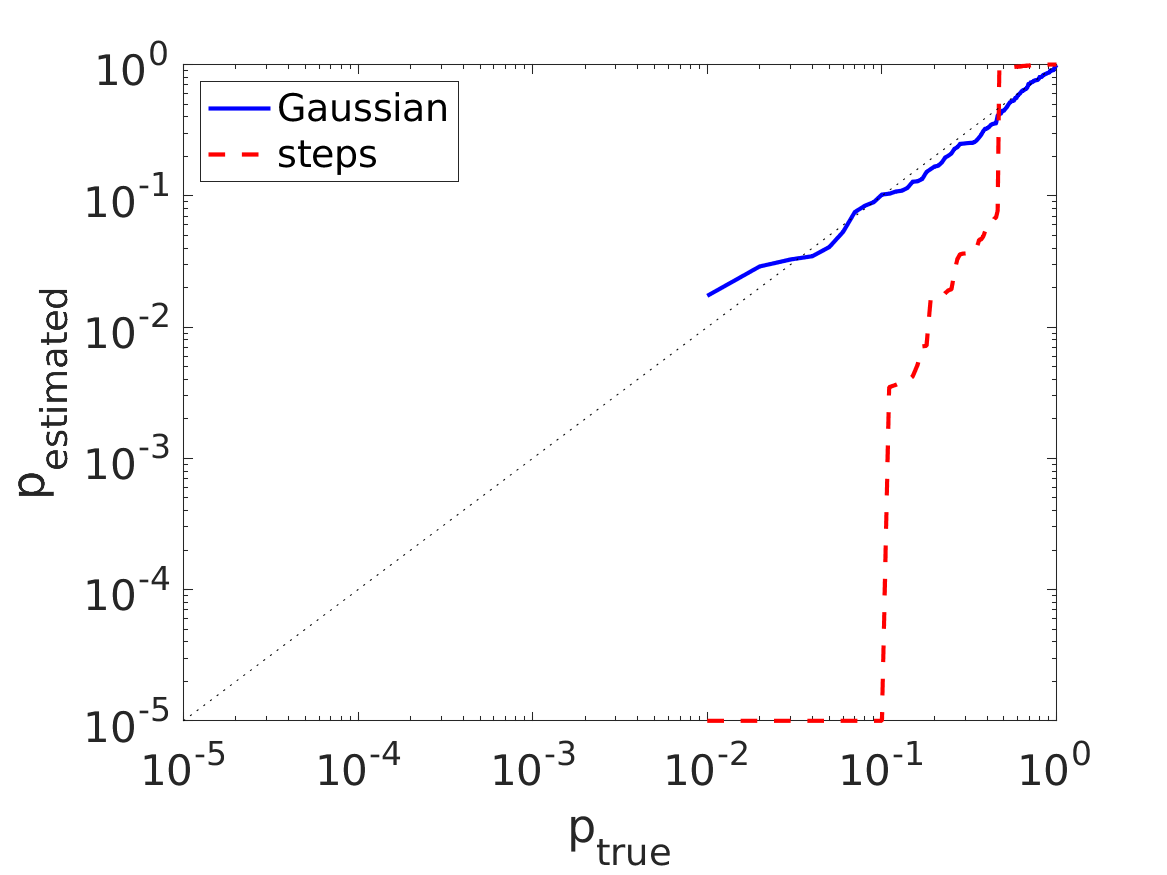}%
    \caption{
    The estimated $p$-value (determined with highly correlated phase scrambles) versus the true $p$-value for a toy-model calculation with two pulsars.
    The dashed red curve shows data consisting of highly non-stationary, non-Gaussian noise.
    This noise is badly misspecified for illustrative purposes, and should not be taken to be representative of realistic pulsar timing noise.
    The solid blue curve shows data consisting of Gaussian noise.
    }
    \label{fig:non-stationary}
\end{figure}

We generate $100$ realizations of noise with five frequency bins and calculate a cross-correlation statistic (Eq.~\ref{eq:rho}).
For each realization, we use the phase scrambling procedure (Eq.~\ref{eq:phase_scramble}) in order to generate $10^5$ highly correlated phase scrambled realizations of the cross-correlation statistic.
We use these phase scrambles to estimate a $p$-value.
In Fig.~\ref{fig:non-stationary} we plot in dashed red this estimated $p$-value against the true $p$-value, which we obtain by simply sorting the 100 unscrambled cross correlation statistic values.
We find that it is relatively common ($\approx 10\%$ probability) to observe small $p$-values $\leq 10^{-5}$.
This is in contrast to Gaussian noise (solid blue), which yields reliable $p$-values.

While contrived---realistic pulsar timing noise models are unlikely to be so badly misspecified---this example illustrates how statistically dependent scrambles can fail due to flawed assumptions. 
Independent scrambles are potentially useful for identifying breakdowns in the assumptions underpinning a scrambling algorithm.
We hypothesize that background estimation is more robust when more independent scrambles are available.

\subsection{Other solutions}
If one wants to test the original null hypothesis, there are options available to obtain additional independent scrambles, which can be employed to probe lower $p$-values.
First, one can increase the number of high-quality measurements by combining data from multiple PTAs \citep[see, e.g.,][]{Perera_2019, Antoniadis_2022} or by timing the pulsars for longer, which increases the number of useful frequency bins.
The more measurements that contribute to the optimal signal-to-noise ratio (Eq.~\ref{eq:rho}), the more independent scrambles become available.
Future IPTA analyses will therefore have the potential for additional scrambles.

Second, one can trade signal-to-noise ratio for an improved understanding of the null distribution.
The pulsar-weighted match statistic (Eq.~\ref{eq:match}) yields more quasi-independent scrambles than the noise-weighted version (Eq.~\ref{eq:sky_match}), which reflects the fact that the least noisy pulsars are weighted as most important in the calculation of $\rho$ (Eq.~\ref{eq:rho}).
However, one may define a ``sub-optimal'' signal-to-noise ratio, which weights pulsars (and/or frequency bins) more evenly.
On average, this reduces the signal-to-noise ratio by assigning sub-optimal weights to noisy measurements.  By making each measurement more comparable, it increases the number of independent scrambles, yielding a better understanding of the background.
For example, one could introduce a new filter function
\begin{align}
    Q_{ij}(f_\mu)^\text{new} = \big( Q_{ij}(f) \big)^\beta ,
\end{align} 
where $\beta \in (0,1)$ controls the shape of the filter.
When $\beta=1$, each measurement is weighted in the traditional way, but when $\beta=0$, each measurement is treated equally.
By varying $\beta$ it may be possible to balance the desire for independent scrambles while still maximizing the signal-to-noise ratio.

Finally, one may eschew scrambles altogether in favour of a detection statistic that is constructed directly from our understanding of gravitational-wave phase measurements.
One may ignore the amplitude of the residuals altogether and construct a statistic entirely using the coherence of different pulsar pairs:
\begin{align}\label{eq:coh}
    \text{coh} = \frac{
    \sum_{i \neq j, \mu}
   \big(\frac{Y_{ij}(f_\mu)}{|Y_{ij}(f_\mu)|}\big)
    \text{sgn}(\Gamma_{ij})
    \,
    \Gamma_{ij}^2 \,
    \sigma_{ij}^{-2}(f_\mu)
    }{
    \sum_{i \neq j, \mu} \Gamma_{ij}^2 \, \sigma_{ij}^{-2} (f_\mu)
    } ,
\end{align}
where
\begin{align}
    Y_{ij}(f_\mu) \equiv s_i^*(f_\mu) s_j(f_\mu) ,
\end{align}
is proportional to the cross-power spectral density and
\begin{align}
    \sigma_{ij}^2(f_\mu) \equiv P_i(f_\mu) P_j(f_\mu) .
\end{align}
This coherence statistic, similar to the one proposed in \cite{Jenet_2005}, ignores entirely information about the amplitude of the observed strain and relies entirely on the phase to determine if a signal is present in the data.
If a gravitational-wave signal is present in the data, the terms in the numerator will tend to be positive, which on average yields a positive value of coh via a biased random walk.
If no signal is present, the coherence tends toward zero.
Ignoring the amplitude information entirely likely leads to some loss of sensitivity.
It is not obvious if a coherence statistic like this is more or less robust than quasi-resampling methods.
Additional work is required to understand the behaviour of the coherence statistic.

\section{Conclusions}\label{discussion}
In order to confidently detect a stochastic gravitational-wave signal, it is necessary for pulsar timing array experiments  to accurately estimate their background.
The problem of background estimation with pulsar timing arrays is interesting as there are subtle differences from the similar problem of background estimation with audio-band observatories like LIGO--Virgo--KAGRA.
While audio-band observatories can generate millions of independent realizations of their background using time slides, current pulsar timing arrays may be limited to $\lesssim 1000$ independent noise realizations through a combination of sky scrambles and phase scrambles.

That said, a large number of independent noise realizations are not necessarily required to calculate a well-defined $p$-value.
One may estimate the probability of obtaining a signal-to-noise ratio at least as large as the observed signal-to-noise ratio among the set of correlated scrambles.
This is not necessarily equivalent to estimating the $p$-value with a set of independent scrambles, but as best we can tell the $p$-value is still well defined.
We hypothesize that background estimation is more robust to erroneous assumptions in the scrambling algorithm when more independent scrambles are available.

In any case, whether one is using independent scrambles or dependent ones, it is desirable to choose the subset of scrambles that are minimally correlated with the unscrambled data.
Otherwise, the background is contaminated by signal, which makes it harder to detect a gravitational-wave signal.
To that end, one should use the appropriate definition of match (Eqs.~\ref{eq:sky_match}, \ref{eq:phase_match}, \ref{eq:super_match}) that takes into account the way the optimal statistic is actually calculated.

It is not clear to us at present the extent to which a PTA experiment should aspire to have a large number of independent scrambles.
Recent work by \cite{Hazboun_2023} shows that the empirical distribution of $\rho$ generated using scrambled NANOGrav data is well matched to the expected theoretical distribution, which provides evidence that the NANOGrav noise models are at least reasonably well specified.
Our intuition is that background estimation is more reliable if it is carried out using a large number of independent noise realizations.
On the other hand, we do not believe that $p$-values calculated from correlated scrambles would \textit{necessarily} yield an excess of false positives, but there may be failure modes in cases where assumptions of stationary Gaussian noise in the timing residuals break down. 

We suggest this topic is worthy of additional consideration.
In particular, we recommend stress-testing pulsar-timing background estimation with simulated noise that has been carefully generated to be subtly misspecified, but plausible.
Posterior predictive checks, such as those proposed in \cite{Vallisneri2023,Myers2023} are another valuable tool for guarding against noise model misspecification.
In the meantime, we suggest methods for generating additional independent scrambles should they prove useful.

\section{Epilogue}\label{epilogue}
Following the submission of this manuscript, several pulsar timing papers appeared presenting varying degrees of evidence in support of a gravitational-wave background detection: \cite{Agazie_2023,Xu_2023,Antoniadis_2023,Reardon_2023}.
In this epilogue we summarize where things stand in light of these results.

The observation of statistically significant signals from NANOGrav, EPTA, and CPTA is encouraging.
The amplitude and spectral index of these three signals appears to be consistent, though, the CPTA signal \citep{Xu_2023} favours higher amplitudes than suggested by NANOGrav and the EPTA \citep{Agazie_2023,Antoniadis_2023}.
The fact that the same signal is not obviously apparent in the PPTA results may be due to the varying sensitivities of each array.
However, the lack of red noise in PPTA observations of PSR J1713+0747 is puzzling as this data is somewhat in tension with the indications of a gravitational-wave background \citep{Reardon_2023}.

A systematic study investigating the consistency between different pulsar timing array datasets will be invaluable.
This would also help to explain aspects of individual datasets that are not entirely well understood, including the evident non-stationarity of the red noise observed by PPTA \citep{Reardon_2023} and the change in the inferred values of $(A,\gamma)$ when EPTA excludes their earliest measurements \citep{Antoniadis_2023}.
Upcoming measurements by MeerKat \citep{bailes_2020,miles_2023} will provide an additional consistency check.

Significant work has been carried out in order to ascertain the reliability of current quasi-resampling methods.
These tests show that the distribution of the optimal statistic is---in practice---not strongly dependent on the choice of threshold for the match statistic, or on the definition of match---whether or not noise weighting is taken into account \citep{Agazie_2023,Antoniadis_2023}.
Moreover, this distribution is also approximately consistent with the theoretical distribution expected if the noise models are well specified \cite{Hazboun_2019}.
All of these developments are extremely promising.

As discussed above, we believe it would be useful to further stress-test this framework using plausibly misspecified noise models.
This would help answer the question: by how much can the $p$-value be shifted due to noise misspecification?
The same study could potentially quantify the extent to which inferences of $(A, \gamma)$ might be affected by subtle noise misspecification.
(This would seem to be especially important in light of the fact that current evidence for stochastic background is inconsistent with the vanilla spectral-index value of $\gamma=13/3$.)
Finally, it is interesting that different scrambling schemes appear to produce different $p$-values: the NANOGrav signal is observed with greater significance using phase scrambles (which emphasise coherence) compared to sky scrambles (which emphasise quadrupolarity).
It wold be helpful to understand the pros and cons of each scrambling method.

\section{Acknowledgements}
We acknowledge and pay respects to the Elders and Traditional Owners of the land on which this work has been performed, the Bunurong, Wadawurrong and
Wurundjeri People of the Kulin Nation and the Wallumedegal People of the Darug Nation.
The work presented here was inspired by the IPTA ``3P+ process,'' by which the IPTA seeks to vet its statements about nanohertz gravitational waves; we thank the members of the IPTA.
We are grateful to fellow members of the IPTA Detection Committee for numerous conversations that helped to shape this work \citep{ipta-dc}: Bruce Allen, Sanjeev Dhurandhar, Yashwant Gupta, Maura McLaughlin, Priya Natarajan, and Alberto Vecchio.
The authors are supported via the Australian Research Council (ARC) Centre of Excellence CE170100004.
E.T. is supported through ARC DP230103088.
V.D.M. receives support from the Australian Government Research Training Program. 
R.M.S. acknowledges support through ARC Future Fellowship FT190100155. 
This work was performed on the OzSTAR national facility at Swinburne University of Technology. The OzSTAR program receives funding in part from the Astronomy National Collaborative Research Infrastructure Strategy (NCRIS) allocation provided by the Australian Government.

\section{Data availability}
The code for this work is available at \url{https://github.com/valeaussie/Sky_scrambles_2}.

\bibliography{refs} 

\appendix

\section{Illustration of misspecification}
\subsection{Overview}
To illustrate the possible effects of model misspecification, we analyze simulated data using noise models that we have intentionally (but subtly) misspecified.
The details of this simulation are provided in Appendix~\ref{simulation}, but we here provide a high-level overview.
We simulate the intrinsic red noise from 26 pulsars in the PPTA array having amplitude $A_N$ and noise consistent with PPTA-DR2. No gravitational wave signal is included in this simulation. We then calculate the cross-correlation and the optimal statistic, systematically underestimating the red noise amplitude by an amount that is plausible based on inference of the PPTA-DR2 data set \cite{Goncharov_2021B}.

\begin{figure}[h]
    \centering
    \includegraphics[width=0.4\textwidth]{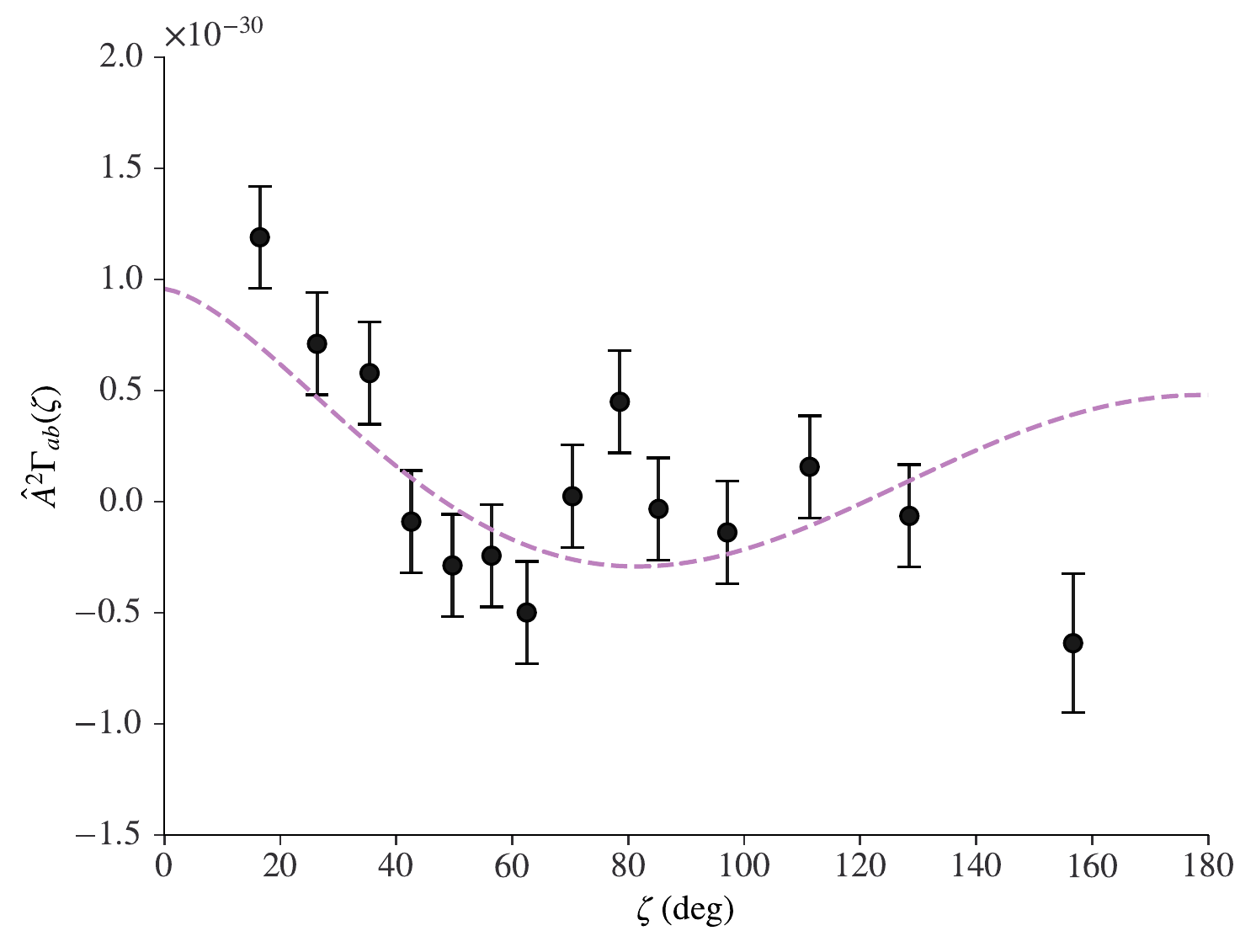}
    \caption{Optimal statistic cross-correlations between pulsar pairs for a simulation containing only noise. Pulsar pairs are grouped into bins and the average cross-correlated power is calculated. In blue the Hellings and Downs curve. The optimal statistic SNR for this realisation was 4.98 and consistent with detection even though no gravitational wave signal was simulated.}
    \label{fig:orf_true}
\end{figure}

We simulate $100$ realizations of pure noise and analyze each one with our misspecified noise model.
Of these $100$ realizations of pure noise, the most significant (as measured with the optimal statistic Eq.~\ref{eq:rho}) yielded $\rho = 4.98$ (nominal $p$-value $=2.26\times 10^{-7}$).
This should be contrasted with what one would expect to be a typical noise fluctuation if the noise were adequately specified: $\rho=2.3$ ($p$-value = 0.01). 
We plot the reconstructed angular correlation function for this simulated data in Fig.~\ref{fig:orf_true}.
The agreement between the Hellings-Downs curve is not perfect, but --- by chance --- a noise fluctuation has produced a quasi-quadrupolar fluctuation that appears to be statistically significant due to model misspecification.
This demonstration emphasizes that an unlucky noise fluctuation can yield a superficially plausible false-positive signal when analyzed with a misspecified noise model.

\subsection{Details of simulation}\label{simulation}

To investigate potential misspecification, applying standard detection statistics to simulated data sets. The simulated data sets are produced with \textsc{PTAsimulate}\footnote{https://bitbucket.org/psrsoft/ptasimulate}. In our case, we use \textsc{PTAsimulate} to generate pulse arrival times using a known pulsar timing ephemeris. We generate 1,000 realisations of 26 PPTA pulsars. All pulsars have the same time of arrival uncertainty of $0.1 \mu s$ and frequency of 1400 MHz and the same intrinsic red noise.

The red noise is parameterised using a power-law power spectral density:
\begin{equation}
    \mathcal{P} = \frac{A_N^2}{12 \pi^2} \bigg( \frac{f}{1 \text{yr}^{-1}} \bigg) ^{- \gamma_N},
\end{equation}
parameterized using an amplitude $A_N$  and spectral index $\gamma_N$. For the simulations, we assume the noise amplitude is  $\log_{10} A_N = -14.46$ and the spectral index $\gamma_N$ is equal to 4.
We simulate data with a constant cadence of $40$\,d over a $22$\,y (8,000 days).
No gravitational waves were injected into the data set.

Here we consider how misspecification affects the \textit{maximum-likelihood} optimal statistic \citep{Anholm_2009, Chamberlin_2015}. This statistic is calculated using the fixed values of the red noise parameters derived by maximising the likelihood function of PTA. We believe similar results will be achieved with the \textit{marginalised} optimal statistic \citep{Vigeland_2018} in which the red noise parameters are drawn from the posteriors generated by the Bayesian analysis of all the pulsars in PTA.

The optimal statistic is calculated using the code in \texttt{enterprise\_extensions}\footnote{https://github.com/nanograv/enterprise\_extensions}.

The plot \ref{fig:orf_true} was obtained by systematically underestimating the log amplitude of the red noise ($\log_{10} A_N$) that \textsc{Enterprise} use for the calculation of the optimal statistic. Specifically, we systematically underestimate the amplitude of the noise for all pulsars, parameterized as 
    \begin{equation}
    \log_{10} A_{\rm model} = \log_{10} A_N - \Delta \log_{10} A.
    \end{equation}

Our simulations highlight how spurious detections can occur if the noise is misspecified.
As expected, when the error bars of the cross-correlations are underestimated due to misspecification, the values of the optimal statistic SNR  are higher than one would anticipate.  Plot \ref{fig:orf_true} shows an instance in which the cross-correlations have a shape that follows the Hellings and Downs curve. In this specific example, the value of the amplitude of the timing noise is underestimated by only $\Delta \log_{10} A_N = 0.15$ (from -14.46 to -14.61). 
This results in the underestimation of the uncertainties of the cross-correlations and in turn, this causes the value of the SNR of the optimal statistic greater than $4 \sigma$ even though we do not include a gravitational wave signal in our simulations. 
In other words, 1\% of the time we observe a $4 \sigma$ SNR, which should occur with 0.003\% frequency.
    
For completeness, we create also three other misspecification scenarios:

\begin{itemize}
    \item We underestimate the log amplitude randomly by drawing the misspecifications from a uniform distribution.
    \item We underestimate and overestimate the log amplitude, again drawing from a uniform distribution, but we put a limit to the overestimation to 1.4\% of the real simulated value of $-14.46$, so we do not expect overestimations higher than $-14.26$.
    \item Then we misspecify the amplitude at random with an equal probability to be overestimated and underestimated.
\end{itemize} 

When underestimating the log amplitude randomly for each pulsar (scenario 2), we find that larger levels of underestimation are required to find significant SNRs. However, one can still get values of the optimal statistic SNR as high as $5 \sigma$ with random underestimations that are not bigger than $\Delta \log_{10} A_N = 0.4$ (from -14.46 to -14.86). We then considered underestimations alongside overestimations of the log amplitude setting an upper limit of $\Delta \log_{10} A_N = 0.2$ on the values of the overestimations. Here again, we see that we can get a $5 \sigma$ optimal statistic SNR with misspecifications not bigger than $\Delta \log_{10} A_N = 0.6$.

As a final remark, we note that values of the optimal statistic SNR higher than expected can be obtained even with equally probable underestimations and overestimations of the $\Delta \log_{10} A_N$. However, the misspecification of the log amplitude, in this case, needs to be higher than $\Delta \log_{10} 0.8$ to find examples where the  optimal statistic SNR is greater than $5 \sigma$.

\section{Derivation of the match statistics}
\subsection{Sky scrambles}\label{app:sky_match}
We calculate the expectation value of the product of two different signal-to-noise ratios, each corresponding to a different scrambled sky:
\begin{align}
    M \equiv \langle \rho \, \rho' \rangle ,
\end{align}
each associated with a different scramble.
This yields

\begin{align} \label{match_ss}
    M = & \frac{
    \Big\langle
    \sum_{i\neq j,\mu} \frac{ s_i^*(f_{\mu}) s_j(f_{\mu}) \, \Gamma_{ij} \, \Omega_\text{gw}(f_{\mu})}
    {f_{\mu}^3 P_i(f_{\mu}) P_j(f_{\mu})}
    \sum_{k\neq l,\beta} \frac{ s_k^*(f_{\beta}) s_l(f_{\beta}) \, \Gamma_{kl}' \, \Omega_\text{gw}(f_{\beta}')}
    {f_{\beta}'^3 P_k(f_{\beta}) P_l(f_{\beta})}
    \Big\rangle
    }{ 
    \left(\sum_{i\neq j,\mu} \, \frac{\Gamma_{ij}^2 \, \Omega_\text{gw}^2(f_{\mu})}
    {f_{\mu}^6 P_i(f_{\mu}) P_j(f_{\mu})} \right)^{1/2}
    \left(\sum_{k\neq l,\beta} \, \frac{\Gamma_{kl}'^2 \, \Omega_\text{gw}^2(f_{\beta}')}
    {f_{\beta}'^6 P_k(f_{\beta}') P_l(f_{\beta}')} \right)^{1/2}
    } \nonumber\\
    = & \frac{
    \sum_{i\neq j,\mu} \frac{\Gamma_{ij} \, \Omega_\text{gw}(f_{\mu})}
    {f_{\mu}^3 P_i(f_{\mu}) P_j(f_{\mu})}
    \sum_{k\neq l,\beta}' \frac{\Gamma_{kl}' \, \Omega_\text{gw}(f_{\beta}')}
    {f_{\beta}'^3 P_k(f) P_l(f_{\beta})}
    \langle s_i^*(f_{\mu}) s_j(f_{\mu}) s_k^*(f_{\beta}') s_l(f_{\beta}') \rangle
    }{ 
    \left(\sum_{i\neq j,\mu} \, \frac{\Gamma_{ij}^2 \, \Omega_\text{gw}^2(f_{\mu})}
    {f_{\mu}^6 P_i(f_{\mu}) P_j(f_{\mu})} \right)^{1/2}
    \left(\sum_{k\neq l,\beta} \, \frac{\Gamma_{kl}'^2 \, \Omega_\text{gw}^2(f_{\beta}')}
    {f_{\beta}'^6 P_k(f_{\beta}') P_l(f_{\beta}')} \right)^{1/2}
    } \nonumber\\
    = & \frac{
    \sum_{i\neq j,\mu} \frac{\Gamma_{ij}\Gamma_{ij}' \, \Omega_\text{gw}^2(f_{\mu})}
    {f_{\mu}^6 P_i(f_{\mu}) P_j(f_{\mu})}
    }{ 
    \left(\sum_{i\neq j,\mu} \, \frac{\Gamma_{ij}^2 \, \Omega_\text{gw}^2(f_{\mu})}
    {f_{\mu}^6 P_i(f_{\mu}) P_j(f_{\mu})} \right)^{1/2}
    \left(\sum_{k\neq l,\beta}\, \frac{\Gamma_{kl}'^2 \, \Omega_\text{gw}^2(f_{\beta}')}
    {f_{\beta}'^6 P_k(f_{\beta}') P_l(f_{\beta}')} \right)^{1/2}
    }.
    \end{align}
In order to obtain the last line, we used the relation
\begin{align}
    \langle s_i^*(f_{\mu}) s_l(f_{\beta}') \rangle = \delta_{il} \, \delta(f_{\mu}-f_{\beta}') \, P_i(f_{\mu}) .
\end{align}
In this derivation, we assume that the noise properties of each pulsar are well characterized so that the power-spectral densities are known.
In practice, they are estimated from the data, and so they are uncertain.
If the signal-to-noise ratio is calculated by marginalizing over power spectral densities, then this calculation becomes more complicated, and the match must incorporate these marginalization integrals as well.
As a first step, however, we opt to use the maximum-likelihood estimators for the power spectral densities.

\subsection{Phase scrambling}\label{app:phase_match}
If we repeat the same exercise with phase scrambling, we set $\Gamma_{ij}=\Gamma_{kl}'$.
However, the strain data pick up random phase constants in every frequency bin. 
We get the following

\begin{align} \label{match_ps}
    M = & \frac{
    \Re \bigg[
    \Big\langle
    \sum_{i\neq j,\mu} \frac{ 
    e^{i(\phi_{j,\mu} - \phi_{i,\mu})} 
    s_i^*(f_\mu) s_j(f_\mu) \, \Gamma_{ij} \, \Omega_\text{gw}(f_\mu)}
    {f_\mu^3 P_i(f_\mu) P_j(f_\mu)}
    \sum_{k\neq l,\beta} \frac{
    e^{i (\phi'_{l,\beta} - \phi'_{k,\beta})}
    s_k^*(f_\beta) s_l(f_\beta) \, \Gamma_{kl}' \, \Omega_\text{gw}(f_\beta)}
    {f_\beta^3 P_k(f_\beta) P_l(f_\beta)}
    \Big\rangle
    \bigg]
    }{ 
    \left(\sum_{i\neq j,\mu} \, \frac{\Gamma_{ij}^2 \, \Omega_\text{gw}^2(f_\mu)}
    {f_\mu^6 P_i(f_\mu) P_j(f_\mu)} \right)^{1/2}
    \left(\sum_{k\neq l,\beta} \, \frac{\Gamma_{kl}^2 \, \Omega_\text{gw}^2(f_\beta)}
    {f_\beta^6 P_k(f_\beta) P_l(f_\beta)} \right)^{1/2}
    } \nonumber\\
    = & \frac{
    \sum_{i\neq j} \Gamma_{ij} \sum_\mu \frac{ \Omega_\text{gw}(f_\mu)}
    {f_\mu^3 P_i(f_\mu) P_j(f_\mu)}
    \sum_{k\neq l} \Gamma_{kl} \sum_\beta \frac{ \Omega_\text{gw}(f_\beta)}
    {f_\beta^3 P_k(f_\beta) P_l(f_\beta)}
    \Re \Big[
    e^{i(\phi_{j,\mu} - \phi_{i,\mu})}
    e^{i (\phi'_{l,\beta} - \phi'_{k,\beta})}
    \langle s_i^*(f_\mu) s_j(f_\mu) s_k^*(f_\beta) s_l(f_\beta) \rangle
    \Big]
    }{ 
    \left(\sum_{i\neq j} \Gamma_{ij}^2 \sum_\mu \frac{\Omega_\text{gw}^2(f_\mu)}
    {f_\mu^6 P_i(f_\mu) P_j(f_\mu)} \right)^{1/2}
    \left(\sum_{k\neq l} \Gamma_{kl}^2 \sum_\beta  \frac{\Omega_\text{gw}^2(f_\beta)}
    {f_\beta^6 P_k(f_\beta) P_l(f_\beta)} \right)^{1/2}
    } \nonumber\\
    = & \frac{
    \sum_{i\neq j} \Gamma_{ij}^2 \sum_\mu \frac{\Omega_\text{gw}^2(f_\mu) \, \Re\Big(e^{i(\phi_{j,\mu} - \phi_{i,\mu} + \phi'_{j,\mu} - \phi'_{i,\mu})}\Big) }
    {f_\mu^6 P_i(f_\mu) P_j(f_\mu)}
    }{ 
    \left(\sum_{i\neq j} \Gamma_{ij}^2 \sum_\mu \frac{\Omega_\text{gw}^2(f_\mu)}
    {f_\mu^6 P_i(f_\mu) P_j(f_\mu)} \right)^{1/2}
    \left(\sum_{k\neq l} \Gamma_{kl}^2 \sum_\beta \frac{ \Omega_\text{gw}^2(f_\beta)}
    {f_\beta^6 P_k(f_\beta) P_l(f_\beta)} \right)^{1/2}
    } \nonumber\\
    = & \frac{
    \sum_{i\neq j} \Gamma_{ij}^2 \sum_\mu \frac{\Omega_\text{gw}^2(f_\mu) }
    {f_\mu^6 P_i(f_\mu) P_j(f_\mu)}
    \cos\big(\phi_{j,\mu}-\phi_{i,\mu} + \phi'_{j,\mu}-\phi'_{i,\mu}\big)
    }{ 
    \left(\sum_{i\neq j} \Gamma_{ij}^2 \sum_\mu \frac{\Omega_\text{gw}^2(f_\mu)}
    {f_\mu^6 P_i(f_\mu) P_j(f_\mu)} \right)^{1/2}
    \left(\sum_{k\neq l} \Gamma_{kl}^2 \sum_\beta \frac{\Omega_\text{gw}^2(f_\beta)}
    {f_\beta^6 P_k(f_\beta) P_l(f_\beta)} \right)^{1/2}
    }
\end{align}
Here, $\phi_{i,\mu}, \phi_{j,\mu}, \phi'_{i,\mu}, \phi'_{j,\mu}$ are the random phases multiplied to the $\mu^\text{th}$ frequency bin of pulsars $i$ and $j$ respectively.
The unprimed $\phi$'s represent phases from one scramble while the primed $\phi$'s represent another scramble.

Our calculations of the sensitivity curves used in the match statistics were performed using \textsc{Hasasia}\footnote{https://github.com/Hazboun6/hasasia} \citep{Hazboun_2019}.

\end{document}